\documentclass[graybox]{svmult}
\usepackage{mathptmx} 
\usepackage{helvet} 
\usepackage{courier} 
\usepackage{type1cm} 
\usepackage{makeidx} 
\usepackage{graphicx} 
\usepackage{multicol} 
\usepackage[bottom]{footmisc}
\makeindex 
\begin{document}
\title*{Stellar Populations in the Galactic Center}
\author{Bono$^{1,2,3}$, G., Matsunaga, N., Inno, L., Lagioia, E.P., and Genovali, K.}
\authorrunning{Bono et al.}
\institute{Bono, G. \at $^1$Dipartimento di Fisica, Universita' di Roma Tor Vergata,
Via della Ricerca Scientifica 1, 00133 Roma, Italy, \email{bono@roma2.infn.it}\\
$^2$INAF-- Osservatorio Astronomico di Roma, via Frascati 33, 00040 Monte Porzio Catone, Italy
$^3$European Souther Observatory, Karl-Schwarzschild-Str. 2 85748 Garching bei Munchen Germany (science visitor)
\and Matsunaga, N. \at Department of Astronomy, School of Science,
The University of Tokyo, 7-3-1 Hongo, Bunkyo-ku, Tokyo 113-0033, Japan,
\email{matsunaga@astron.s.u-tokyo.ac.jp}
\and Inno, L. \at Dipartimento di Fisica, Universita' di Roma Tor Vergata,
Via della Ricerca Scientifica 1, 00133 Roma, Italy, \email{laura.inno@roma2.infn.it}
\and Lagioia, E. \at Dipartimento di Fisica, Universita' di Roma Tor Vergata,
Via della Ricerca Scientifica 1, 00133 Roma, Italy, \email{eplagioia@roma2.infn.it}
\and Genovali, K. \at Dipartimento di Fisica, Universita' di Roma Tor Vergata,
Via della Ricerca Scientifica 1, 00133 Roma, Italy, \email{katia.genovali@roma2.infn.it}
} 
\maketitle
\abstract*{We discuss the stellar content of the Galactic Center, and in particular,
recent estimates of the star formation rate (SFR). We discuss pros and cons
of the different stellar tracers and focus our attention on the SFR based on
the three classical Cepheids recently discovered in the Galactic Center.
We also discuss stellar populations in field and cluster stars and present
some preliminary results based on near-infrared photometry of a field centered
on the young massive cluster Arches. We also provide a new estimate of the
true distance modulus to the Galactic Center and we found
14.49$\pm$0.02(standard)$\pm$0.10(systematic) mag (7.91$\pm$0.08$\pm0.40$ kpc).
Current estimate agrees quite well with similar photometric and kinematic distance
determinations available in the literature. We also discuss the metallicity
gradient of the thin disk and the sharp change in the slope when moving
across the edge of the inner disk, the Galactic Bar and the Galactic Center.
The difference becomes even more compelling if we take into account that
metal abundances are based on young stellar tracers (classical Cepheids,
Red Supergiants, Luminous Blue Variables). Finally, we briefly outline the
possible mechanisms that might account for current empirical evidence.
}
\abstract{We discuss the stellar content of the Galactic Center, and in particular,
recent estimates of the star formation rate (SFR). We discuss pros and cons
of the different stellar tracers and focus our attention on the SFR based on
the three classical Cepheids recently discovered in the Galactic Center.
We also discuss stellar populations in field and cluster stars and present
some preliminary results based on near-infrared photometry of a field centered
on the young massive cluster Arches. We also provide a new estimate of the
true distance modulus to the Galactic Center and we found
14.49$\pm$0.02(standard)$\pm$0.10(systematic) mag (7.91$\pm$0.08$\pm0.40$ kpc).
Current estimate agrees quite well with similar photometric and kinematic distance
determinations available in the literature. We also discuss the metallicity
gradient of the thin disk and the sharp change in the slope when moving
across the edge of the inner disk, the Galactic Bar and the Galactic Center.
The difference becomes even more compelling if we take into account that
metal abundances are based on young stellar tracers (classical Cepheids,
Red Supergiants, Luminous Blue Variables). Finally, we briefly outline the
possible mechanisms that might account for current empirical evidence. 
} 
\section{Introduction} \label{sec:1}
The Galactic Bulge and the Galactic Center play a crucial role in
constraining the formation and the evolution of the Galactic spheroid.
Recent numerical simulations indicate that the Milky Way formed Inside-Out,
this means that the bulge harbors the oldest Galactic populations
~\cite{debattista06}. This
theoretical framework is soundly supported by recent photometric
investigations ~\cite{zoccali03} suggesting that stellar populations
in the Galactic Bulge are mainly old ($\sim$11 Gyr).
On the other hand, the Galactic Center together with the super-massive black hole,
harbors, very young stars (a few Myr, ~\cite{serabyn96, figer04}), compact
star clusters and massive molecular clouds ~\cite{launhardt02}.

Although current knowledge of the innermost components of the Galactic spheroid
is quite solid, we still lack quantitative constrains concerning their kinematic
structure and their chemical enrichment history. In particular, we still lack
firm estimates of the edge between the thin disk and the Galactic Bulge.
Moreover and even more importantly, current predictions suggest that a presence
of a bar-like structure is crucial to support the high rate of star formation
present in the Nuclear Bulge. Indeed, it is the bar-like structure to
drag the gas and the molecular clouds from the inner disk into the Nuclear Bulge
~\cite{athanassoula92, kim11a,kim11b}.

A more quantitative understanding of this phenomenon will have an impact
into the formation and the evolution of classical bulges and pseudo-bulges.
The latter are considered disk-like stellar components slowly evolving
in galaxy centers, whereas the former are considered the aftermath of
galaxy mergers ~\cite{kormendy04}.
The chemical evolution of the thin disk in the solar circle has been
investigated using different stellar tracers and optical spectroscopy
(Cepheids: ~\cite{lemasle08,pedicelli09}, open cluster: ~\cite{carraro10}). 
The same outcome applies for the Galactic Bulge
~\cite{barbuy09,zoccali08}. The chemical enrichment
in the Galactic Center is still in its infancy ~\cite{najarro09,davies09a}. 

In the following we discuss recent findings concerning stellar populations 
in the Galactic Center. In particular, we will focus our attention on 
recent estimates of the star formation rate (\S2) and on recent findings 
concerning field and cluster stellar populations (\S3). Moreover, we will 
discuss in \S4 recent determinations of the distance to both the Galactic Center and 
the Galactic Bulge. The metallicity distribution will be discussed in \S5 while in \S6 
we briefly outline future perspective for photometric and spectroscopic 
investigations. 
\section{Stellar populations and star formation in the Galactic Center}
\subsection{Star formation}\label{subsec:3} 
The star formation rate (SFR) across the Galactic Center was estimated by ~\cite{genzel97}
and more recently by~\cite{yusef09}, they found that it was constant during
the last few tens of Myr. The latter investigation is based on a paramount analysis
of multiband data: NIR (2MASS), MIR (SPITZER), submillimeter (SCUBA) and radio (VLA).
These data were used to fit the spectral energy distribution of Young Stellar Objects
and on the basis of their ages they provided solid constraints on the recent star
formation rate.
On the other hand, ~\cite{vanloon03}, by using NIR (DENIS) and MIR (ISOGAL)
observations and Asymptotic Giant Branch stars as stellar tracers, found that the
SFR was enhanced in the nuclear bulge more than 200 Myr ago and almost continuous
till present times. By using deep NIR color-magnitude diagrams based on NICMOS
at HST, ~\cite{figer04} found that the present day enclosed mass within
30 pc of the Galactic Center, the star counts and the shape of the K-band luminosity
function support a continuous star formation history with a rate of
$\sim$0.02 $M_\odot$ yr$^{-1}$. This finding is also consistent with the presence
of the three young massive clusters located inside the central 50 pc.
The above investigations are somehow hampered by uncertainties affecting
distance determinations and by the relevant changes of the extinction
across the Galactic Center. Moreover, both individual and ensemble age
estimates based on the above tracers might be affected by possible
systematic uncertainties. Classical Cepheids present several advantages
when compared with other stellar tracers.
\begin{figure}
\centering
\includegraphics[height=0.50\textheight,width=0.99\textwidth]{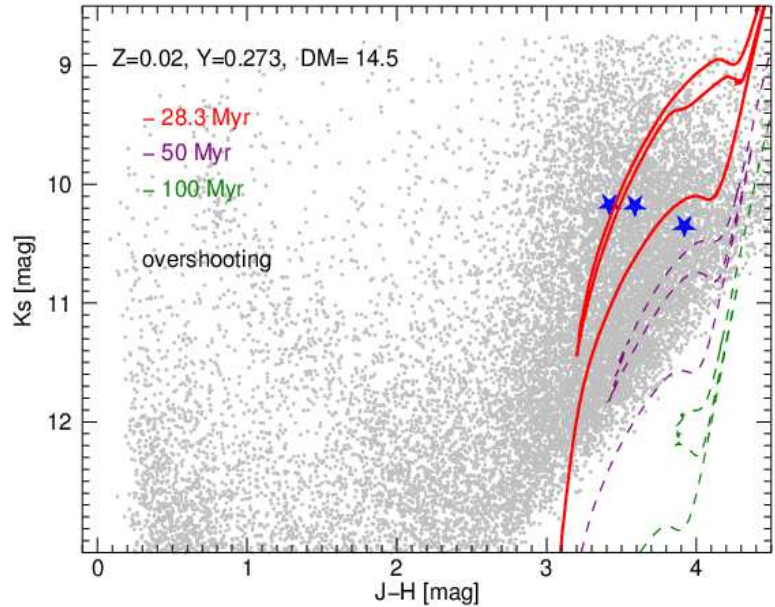}
\caption{Comparison between the NIR photometry of the Galactic Center,
collected with the NIR camera SIRIUS at the 1.4m telescope IRSF in
South Africa, and stellar isochrones. The blue stars mark the position
of the three classical Cepheids recently discovered by
~\cite{matsunaga11}. Solid and dashed lines display isochrones
with ages ranging from 28.3 Myr (red) to 100 Myr (green) and solar
chemical composition (BASTI data base). Isochrones plotted in this
figure account for a mild convective core overshooting during central
hydrogen burning phases. Predictions were plotted by assuming a true
distance modulus of 14.5, an extinction $A_K=2.5$ mag and the reddening
law by ~\cite{nishiyama09}.
}
\label{scheme}
\end{figure}
\begin{enumerate}
\item{{\bf Easy targets}-- Classical Cepheids are bright stars and they can be
easily recognized by using either the Bailey diagram (luminosity amplitude vs period)
or the Fourier parameters of light curves.
}

\item{{\bf Distance indicators}-- Classical Cepheids are very robust primary distance
indicators. NIR Period-Luminosity relations and Period-Wesenheit relations
are marginally affected by metallicity effects and by differential reddening.
}

\item{{\bf Evolutionary status}-- Their evolutionary status is well established. They
are intermediate-mass stars during central helium burning and hydrogen shell burning.
They are evolving along the so-called blue loop and obey to well defined Period-Age
and Period-Color-Age relations ~\cite{bono05}. In particular, once the
chemical composition has been fixed and we neglect the width in temperature of the
instability strip, longer is the pulsation period younger is the age of the Cepheid.
}

\item{{\bf Pulsation properties}-- Their pulsation behavior is also well established
~\cite{bono99}. They pulsate in the fundamental, first overtone and as mixed
pulsators (topology of the Cepheid instability strip). They obey to well defined
Period-Color relations. This means that their apparent colors can be adopted to
provide an independent estimate of the reddening.
}
\end{enumerate}

However, they are far from being ideal stellar tracers, since they also have some
indisputable disadvantages.
\begin{enumerate}
\item{{\bf Time series}-- The identification does require homogeneous time series
data over time intervals ranging from a few days to more than one hundred days.
}

\item{{\bf Multiband observations}-- The use of their mean colors does require
well sampled light curves in at least two different bands.
}

\item{{\bf Pulsation amplitude}-- Their pulsation amplitude steadily decreases
when moving from the optical to the NIR bands. The difference is mainly caused
by the fact that optical bands are more sensitive to temperature variations,
while the NIR bands to radius variations.
}
\end{enumerate}

On the basis of the above circumstantial evidence and on the discovery of three
classical Cepheids with pulsation periods of the order of 20 days, located within
40 parsecs of the central black hole, ~\cite{matsunaga11} provided a solid
estimate of the SFR. On the basis of the lack of classical Cepheids with shorter
periods they estimated that approximately 25 million years ago the star
formation rate in this region increased relative to the period of 30-70 million
years ago.
In order to further constrain the age of the newly discovered Galactic Center Cepheids, we
decide to use an independent approach. Fig.~1 shows the NIR ($K_{\rm S}$ vs $J-H$) Color-Magnitude
Diagram of the Galactic Center region based on NIR images collected with SIRIUS at the 1.4m telescope
IRIS in South Africa. 

To constrain the age of the Cepheids we adopted the stellar
isochrones available in the BaSTI data base\footnote{The interested reader is referred to
http://albione.oa-teramo.inaf.it/} ~\cite{pietrinferni04,pietrinferni06}. In particular, we selected
isochrones at solar chemical composition (global metallicity, Z=0.02; primordial
helium, Y=0.273) accounting for mild convective core overshooting during central
hydrogen burning phases and for a canonical mass loss rate. The ages range from
28 (solid red) to 100 (dashed green) Myr. We also adopted a distance modulus of
$\mu$=14.5 mag (see section 3) and the K-band extinction suggested by
~\cite{matsunaga11} and the reddening law by ~\cite{nishiyama06}. 
The comparison between stellar isochrones and observed classical Cepheids 
(blue stars) indicate that their age is of the order of 28 Myr. This estimate 
agrees quite well with the age based on the Period-Age relation (20--30 Myr). 
It is worth mentioning that age estimates based on the comparison in the Color-Magnitude
diagram (CMD) between data and isochrones do depend on uncertainties distance and reddening 
estimates. Moreover, they are affected by uncertainties in the treatment of 
mass loss rate and in the efficiency of internal mixing 
~\cite{pradamoroni12}. The age estimates based on the Period-Age 
relations of classical Cepheids are also affected by uncertainties 
affecting the mass-luminosity relation, but they appear to be more 
robust concerning individual age determinations ~\cite{bono05}. 

\subsection{Stellar populations}\label{subsec:2}
In two seminal investigations concerning the stellar populations in the Galactic Center,
dating back to more than twenty years ago, ~\cite{philipp99} and ~\cite{mezger99}
performed a NIR survey (the pixel scale was 0.278 arcsec and the filed of view
$\approx$71$\times$71 arcsec) of the central 30 pc using IRAC at 2.2m ESO/MPG telescope.
They provided firm empirical constraints on the role played by the different stellar
components. They found that low-mass (M$< 1M_\odot$) stars are the main contributor to
the dynamical mass (90\%), while they only contribute a minor fraction of the K-band
flux, namely the 6\%. On the other hand, intermediate- and high-mass stars contribute
with only the 6\% of the dynamical mass, but with almost the 90\% of the K-band flux.
By using NIR, MIR, millimeter (IRAM) and radio (MPIfR) observations, ~\cite{launhardt02} confirmed the previous
finding, and indeed they found that 70\% of the optical-UV flux comes from massive
stars. Moreover, they also provided an estimate of the dynamical mass and thy found
that it is M$\approx$1.4$\times$10$^9$ $M_\odot$.
\begin{figure}
\centering
\includegraphics[height=0.65\textheight,width=0.99\textwidth]{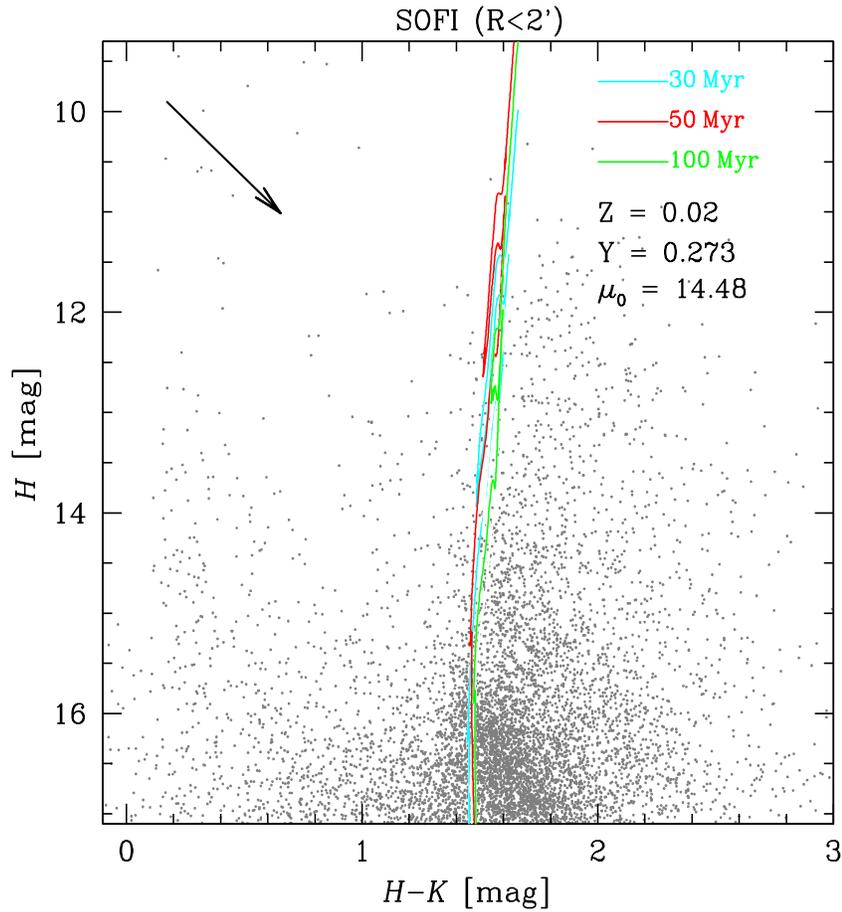}
\caption{Comparison between the NIR photometry of the young cluster
Arches collected with the NIR camera SOFI at NTT/ESO in La Silla,
and stellar isochrones. Note that are only plotted stars located within two
arcminutes from the cluster center. The solid lines show young 
(30 Myr) isochrones at solar chemical composition (BaSTI data base).
The adopted distance modulus of 14.5, the K-band extinction $A_K=2.5$ mag 
and the reddening law are the same as in Fig.~1. The black arrow shows 
the reddening vector. 
}
\label{scheme}
\end{figure}

The recent literature concerning the young massive star clusters in the Galactic Center is even more
flourishing. Detailed spectroscopic investigations of high-mass stars in the Central
Cluster and in Arches have been provided by using the integral field spectrograph SINFONI
at ESO/VLT ~\cite{martins07,martins08}. A similar analysis was recently performed by
~\cite{liermann10} for Wolf--Rayet stars along the nitrogen sequence centrally
located in the Quintuplet cluster. They found that their sample is quite bright, but with
relatively cool effective temperatures. Interestingly enough, they
also found that these objects together with similar objects in Arches and in the
Galactic Center form a distinct group in the Hertzsprung-Russell diagram. 
Deep NIR CMDs for Arches and Quintuplet have been
provided by ~\cite{figer99,figer02} using NICMOS at HST.
They reached a limiting magnitudes
between 19 and 20.5 in the F160W and F205W bands and estimated that the cluster age
ranges from 2$\pm$1 Myr (Arches) to 4$\pm$1 Myr (Quintuplet).

A more detailed NIR photometric analysis of the Arches stellar content was recently
provided by  ~\cite{espinoza10}. They adopted NACO Adaptive Optic camera at
ESO/VLT to perform very accurate and deep photometry, they found that a Salpeter--like
power law cannot be discarded for the Initial Mass Function (IMF) of Arches.
This is a very interesting finding, since previous investigations
~\cite{stolte05,kim06} suggested a top--heavy IMF for this cluster.
The empirical scenario is becoming even more interesting due to the ongoing discoveries
of new clusters. By using images collected with NICMOS at HST and low-resolution
spectra with ISAAC at ESO/VLT, ~\cite{davies12} identified a new young massive
cluster located at the far end of the Galactic Bar.

In order to constrain the field and the cluster stellar populations in the Galactic Center, 
we started a long term project aimed at providing a detailed census of the young-, 
the intermediate- and the old populations. We are facing two important problems in 
dealing with a quantitative analysis of the stellar populations in the Galactic Center: 
a) heavy differential extinction affecting
both field and cluster stars; b) transformations into a standard photometric system becomes
a difficult problem in presence of strong reddening variations ~\cite{espinoza10}.
To properly address the latter problem, we performed accurate NIR photometry 
across the Arches cluster by using a set of $J$ (82), $H$ (45) and $K_{\rm S}$ (45) 
images collected with SOFI at ESO/NTT. The entire data set includes both low- 
(0.288 arcsec/px, FoV = 4.9 x 4.9 arcmin) and high- (0.144 arcsec/px, 
FoV = 2.5 x 2.5arcmin) resolution mode. The photometry was performed by using 
DAOPHOT and ALLFRAME (~\cite{stetson94}, and references therein). 
The final photometric catalog was calibrated into the 
2MASS photometric system by using local standards ~\cite{lagioia12}.
Fig.~2 shows a preliminary $H$, $H-K$ CMD of the cluster region located 
within two arcmin from the center. 

The comparison between theory and observations was performed by using 
young cluster isochrones at solar chemical composition. 
The solid lines show isochrones with ages ranging from 30 to 100 Myr. 
The isochrones were computed using evolutionary tracks that account 
for mild convective core overshooting. 
The isochrones were plotted by adopting a true distance modulus of 
14.48$\pm$0.13 mag, an extinction in the K-band of $A(K_{\rm S})$=2.5 mag 
and the reddening law by ~\cite{nishiyama06}. 
A glance at the data plotted in this figure shows that the blue stars 
located at $H-K_{\rm S}\sim$1.4 are cluster young main sequence stars 
(see Fig.~4 in ~\cite{figer09}). 
However, a detailed comparison between theory and observations does 
require an appropriate subtraction of field stars by using nearby 
control fields.
\section{The distance to the Galactic Center} \label{sec:2}
In the recent literature distance estimates to the Galactic Center are often
mixed with distance determinations to the Galactic Bulge. However, current
evaluations concerning the edge of the inner disk suggest d$\sim$5$\pm$1 kpc.
Photometric and astrometric distances to the Galactic Center have been
recently reviewed by ~\cite{matsunaga12}[and references therein]. They
found that both photometric and kinematic estimates cluster around 8kpc.
This means that distance determinations to the Galactic Bulge do depend
on the density profile and on the radial distribution of the adopted
standard candle. It is plausible to assume that the two sets of distance
determinations may differ at the 20\% level. 

In order to provide a
quantitative estimate of the difference we decided to provide a new distance
determination to the Galactic Center by using NIR and MIR Period-Wesenheit
relations for the three new classical Cepheids in the Galactic Center.
The use of the PW relation has several indisputable advantages when
compared with Period-Luminosity relations~\cite{bono10}.

\begin{enumerate}
\item{{\bf Reddening uncertainty}-- The PW relations are independent 
of uncertainties affecting reddening estimates.} 

\item{{\bf Instability strip topology}-- The PW relations are independent 
of the pulsator distribution inside the instability strip, since they 
account for the width in temperature.}

\item{{\bf Linearity }-- They are almost linear over the entire period 
range~\cite{inno12}, since they mimic a PLC relation.}

\item{{\bf Mixing length}-- The dependence of the PW relations (slope and zero-point) 
on the adopted mixing-length parameter is negligible.}

\item{{\bf Chemical composition}-- The slopes of both NIR and optical-NIR PW relations 
appear to be independent of chemical composition for metallicities ranging from Z=0.004 
(Small Magellanic Cloud) to Z=0.02 (Milky Way). The adopted helium content Y, at fixed 
metallicity and mass-luminosity relation, only affects the zero-point of the 
PW relations.}
\end{enumerate}

However, the PW relations also have some indisputable disadvantages.

\begin{enumerate}
\item{{\bf Reddening law}-- The Wesenheit magnitudes rely on the assumptions that the 
reddening law is universal ~\cite{inno12b}. Thus, distance estimates based on 
PW relations do depend on the reddening law adopted to estimate the extinction 
coefficients.}

\item{{\bf Multiband photometry}-- The PW relations require time series data in two 
different bands.}
\end{enumerate}

\begin{table}
\caption{Nuclear Bulge Cepheid Distances}
\label{tab:1} 
\begin{tabular}{p{2cm}p{2.4cm}p{2.4cm}p{2.4cm}p{2.4cm}}
\hline\noalign{\smallskip}
Star&$\mu_{WJK_{\rm{s}}}^{a}$& $\mu_{WJH}^{a}$&$\mu_{WHK_{\rm{s}}}^{a}$&$\mu_{mean}^{b}$\\
\noalign{\smallskip}\svhline\noalign{\smallskip}
a &14.69 $\pm$ 0.05 &14.55 $\pm$ 0.05 &14.53 $\pm$ 0.05 &14.57 $\pm$ 0.03 \\
b & 14.51 $\pm$ 0.05 &14.32 $\pm$ 0.05 &14.30 $\pm$ 0.05 & 14.46 $\pm$ 0.03 \\
c & 14.53 $\pm$ 0.05 &14.48 $\pm$ 0.05 &14.41 $\pm$ 0.05 &14.42 $\pm$ 0.03 \\
\noalign{\smallskip}\hline\noalign{\smallskip}
\end{tabular}
$^a$ Distance modulus based on the zero-point calibration obtained by the predicted
FU PW relations for Galactic Cepheids provided by ~\cite{marconi05}. The associated 
error is the standard deviation from the theoretical PW relation. The color 
coefficients of the adopted PW relations are the following:
$\frac{A_K}{ E(J-K_{\rm S})}$=0.50;
$\frac{A_H}{E(J-H)}$ =1.42;
$\frac{A_K}{E(H-K_{\rm S})}$ =1.44;
~\cite{nishiyama06}. \\
$^b$ The weighted average of the three distance modulus estimations.
\end{table}

The distance of the Cepheids in the Galactic Center are derived using the PW relations
in the three NIR bands ($JHK_{\rm S}$). We have defined the three Wesenheit 
magnitudes adopting the color coefficients given in Tab.~\ref{tab:1}.
By using the theoretical models provided by ~\cite{marconi05}, we have computed the
following PW relations:
$$W(JK_{\rm{s}})= - (2.802 \pm 0.002) - (3.205 \pm 0.002) \times \log P $$ 
$$W(JH)=- (2.971 \pm 0.00) - (3.361 \pm 0.007) \times \log P $$ 
$$W(HK_{\rm{s}})=- (2.714\pm 0.003) - (3.123 \pm 0.003) \times \log P $$ 
\noindent 
with standard deviations of 0.03, 0.01 and 0.04 mag.

The mean $JHK_{\rm S}$ magnitudes of the Cepheids are given in
~\cite{matsunaga11}. The difference between the predicted and 
the observed value of the Wesenheit magnitude gives the true 
distance modulus of each Cepheid.
The distance moduli obtained in each band and the error-weighted average
are listed in Tab.~\ref{tab:1}.
The associated error is due to the standard deviation from the theoretical
PW relation and also to the error on the total-to-selective extinction
ratio, as given in ~\cite{nishiyama06}.
The derived distances agree quite well with the independent determinations 
provided by ~\cite{matsunaga11}.
Our results are independent of the extinction correction, therefore, 
we obtain a mean distance modulus for the three Cepheids in the Galactic 
Center of 14.49$\pm$0.02(standard)$\pm$0.10(systematic) mag with a small 
intrinsic error. The systematic error accounts for uncertainties in the 
zero-point of the NIR PW relations and in the reddening law ~\cite{inno12}.  
Thus, we found a mean distance of 7.91$\pm$0.08$\pm0.40$ kpc to the Galactic 
Center that is in very good agreement with the distance to the Galactic Center 
based on the S2 orbit around the central black hole 
(8.28 $\pm$ 0.35 kpc, ~\cite{gillessen09}) and with the parallax 
of Sgr B (7.9 $\pm$ 0.8 kpc, ~\cite{reid09b}).

\section{Metallicity distribution in the Galactic Center}\label{sec:4}
The iron and the $\alpha$-element abundance gradients across the Galactic disk
are fundamental observables to constrain the chemical enrichment of disk stellar
populations ~\cite{andrievsky04,luck06,lemasle08,pedicelli09,luck11a,luck11b}.
They also play a key role in constraining the physical assumptions adopted in
chemical evolution models ~\cite{portinari00,chiappini01,cescutti07}.
The most recent theoretical and empirical investigations
brought forward three open issues:

\begin{enumerate}
\item{{\bf Stellar tracers}-- Empirical evidence indicates that different
stellar tracers do provide different slopes. Metallicity gradients based
on Cepheids, provide slopes ranging from $-0.05$~dex~kpc$^{-1}$
~\cite{luck03,caputo01,lemasle07,kovtyukh05,andrievsky02,luck06,yong06}
to $-0.07$~dex~kpc$^{-1}$ ~\cite{lemasle08}. More recently,
~\cite{pedicelli09} using iron abundances for 265 classical Cepheids
--based either on high-resolution spectra or on photometric metallicity indices--
and Galactocentric distances ranging from $R_G \sim$5 to $R_G \sim$17 kpc, found
an iron gradient of $-0.051\pm0.004$~dex~kpc$^-1$. By using an even more
large sample of over 400 classical Cepheids, ~\cite{luck11b} found a
gradient of $-0.062\pm 0.002$~dex~kpc$^{-1}$.
A similar gradient was also found by ~\cite{friel02} by using a sample of
40 open clusters located between the solar circle and $R_G \sim14$ kpc,
namely $-0.06$~dex~kpc$^{-1}$. On the other hand,~\cite{carraro07} by using
new metallicities for five old open clusters located in the outer disk
(12$\le R_g \le$21 kpc) and the sample adopted by ~\cite{friel02}
found a shallower iron gradient: $-0.018$~dex~kpc$^{-1}$.
The slope of the metallicity gradient based on oxygen abundances of HII regions
--with Galactocentric distances ranging from 5 to 15 kpc-- is similar to the
slope based on Cepheids ($-0.04$~dex~kpc$^{-1}$,~\cite{deharveng00}).
The difference between the different tracers, might be due to the age difference
between the different tracers (young vs intermediate-age).
}

\item{{\bf Linear slope}-- By using a sample of 76 open clusters with distances
ranging from 6 to 15 kpc, it was suggested by ~\cite{twarog97} that a proper
fit to the metallicity distribution does require two zones. The inner disk for
Galactocentric distances ranging from 6 to 10 kpc and the outer disk for distances
larger than 10 kpc. This hypothesis was supported by ~\cite{caputo01,luck03,andrievsky04}.
More recently, ~\cite{pedicelli09} found for the two zones a slope of
$-0.130\pm0.015$~dex~kpc$^{-1}$ for the inner disk ($R_G <$8 kpc) and a slope of
$-0.042\pm0.004$~dex~kpc$^{-1}$ for the outer disk.
Data plotted in Fig.~3 support the above results. We adopted the same abundances
provided by ~\cite{pedicelli09}, but we used new individual distances based on
NIR Period-Wesenheit relations ~\cite{inno12} instead of NIR Period-Luminosity
relations.
Current findings indicate that young tracers do show evidence of a sharp steepening
of the slope in the inner disk and a mild flattening of the gradient in the outer
disk.
}

\item{{\bf Local inhomogeneities}-- There is mounting empirical evidence that
abundance inhomogeneities are present not only across Galactic quadrants, but also
on smaller spatial scales. ~\cite{pedicelli09} found that the iron
abundance of the Cepheids belonging to the two overdensities located in the
second and in the fourth quadrant, covers a range in metallicity similar
to the range in metallicity covered by the global gradient (see also
~\cite{luck06,lemasle07,luck11a}).
}
\end{enumerate}

Fig.~3 shows the position of bright supergiants in two Galactic Center young clusters
--Arches and Quintuplet-- by ~\cite{najarro04,najarro09} (cross) and Galactic Center field supergiants
by ~\cite{davies09a} (open square). The open diamonds display the position of the two Scutum
Red Supergiant clusters by ~\cite{davies09b}) located at the end of the
Galactic Bar. A glance at the data plotted in this figure discloses that
stars located in the Galactic Center have a solar iron abundance, whereas the 
average iron abundances are subsolar by 0.2--0.3 dex for the stars located along 
the Galactic Bar . This means that we are facing a stark discrepancy between the
metallicity gradient based on classical Cepheids and the above iron 
abundances\footnote{The interested reader is referred to the Table~2 of 
~\cite{davies09a} for a detailed list of the most recent metallicity 
estimates of the Galactic Center.}. 
Plain leading arguments based on the extrapolation of both the global and the
inner disk metallicity gradient would imply a super-solar iron abundances
approaching the Galactic Center~\cite{davies09a}). This discrepancy has
already been noted in the literature and ~\cite{cunha07} suggested that
the slope of the metallicity gradient should become shallower for Galactocentric
distances smaller than 5~kpc. In principle there is no plausible reason why
it should be extrapolated from the inner disk to the Galactic Center. Solid empirical
evidence suggest that the Galactic Bulge is dominated by an old (11 Gyr) stellar
population with a secondary intermediate-age (a few Gyr) stellar component
~\cite{zoccali03}. Moreover and even more importantly, recent spectroscopic
investigations based on high-resolution spectra indicates that the stellar populations
in the Galactic Bulge range from metal-intermediate ([Fe/H]$\sim$-1) to super solar  
([Fe/H]$\sim$+0.5) with a broad main peak [Fe/H]$\approx$0.2
~\cite{zoccali03,zoccali08,hill11}.

\subsection{Transition between the Galactic Bulge and the Galactic Center}\label{subsec:4} 
Quantitative constraints concerning the edge between the inner disk and the
Bulge and the edge between the Bulge and the Galactic Center are hampered
by lack of accurate distance determinations (see Fig.~4). The above scenario 
is further complicated by the possible presence of the Galactic Bar. We are 
facing the evidence that the region located between the Galactic Center
and the inner disk is characterized by the lack of star forming regions,
of giant HII regions and of young open clusters ~\cite{davies09a}.
On the other hand, Galactic Legacy Mid-Plane Survey Extraordinaire (GLIMPSE)
survey based on 30 million mid-infrared (SPITZER) sources identified a linear
bar passing through the Galactic Center with half-length $R_{bar}$=4.4$\pm$ 0.5 kpc.

Theoretical ~\cite{athanassoula92, friedli95} and observational
~\cite{zaritsky94, allard06, zanmar08} investigations indicate that
the abundance gradient in barred galaxies
is shallower than in unbarred galaxies. The typical explanation for this
trend is that the bar is dragging gas from the inner disk into the Galactic
Center ~\cite{kim11a}. The pileup of the new fresh material triggers
an ongoing star formation activity till the dynamical stability of the bar.
In passing we note that the three new Cepheids discovered by ~\cite{matsunaga11} 
in the inner disk is supporting the evidence that star formation events might 
have occurred on an area broader than the near end of the Galactic Bar. 
A more quantitative understanding of this phenomenon has an impact
into the formation and the evolution of classical bulges and pseudo-bulges.
The latter are considered disk-like stellar components slowly evolving
in galaxy centers, whereas the former are considered the aftermath of
galaxy mergers ~\cite{kormendy04,kormendy09,matsunaga11}.

\begin{figure}
\centering
\includegraphics[height=0.50\textheight,width=0.99\textwidth]{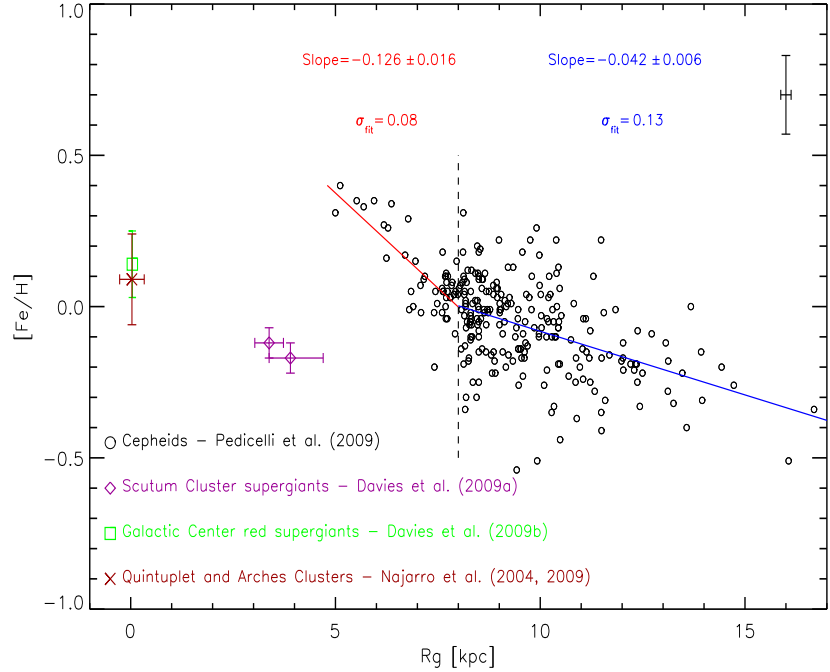}
\caption{Galactic iron gradient as a function of the Galactocentric distance.
Empty circles mark the Galactic Cepheid sample by ~\cite{pedicelli09}.
The solid lines display the metallicity gradient for Cepheids located either
inside (red, 50 objects) or outside (blue, 215 objects) the solar circle
(RG $<$ 8 kpc, vertical dashed line). The slopes and the standard deviations
of the linear fits are also labeled. The error bars in the top right corner
shows the typical uncertainties in iron abundance and in distance.
The two purple diamonds display the metallicity of two supergiants located
in the Scutum cluster ~\cite{davies09a}, while the brow cross the
metallicity of the young cluster Quintuplet ~\cite{najarro09}
and the green square the metallicity of the red supergiants observed by
~\cite{davies09b} in the Galactic Center.}
\label{scheme}
\end{figure}

One of the key consequences of the above scenario is that the typical metallicity
distribution along the Galactic Bar and in the Galactic Center should be
quite similar to the metallicity distribution in the inner disk. The main
advantage in current analysis is that we are using stellar tracers with
similar ages, since both Supergiants, Luminous Blue Variables (LBVs) and
Cepheids are either massive or intermediate--mass stars. Their evolutionary
lifetime is typically shorter than 100 Myr. This evidence seems to support
the hypothesis that the occurrence of the bar might not be the main culprit
in shaping the metallicity gradient between the inner disk and the Galactic
Center. Indeed, current numerical simulations suggest that the timescale
within which the radial motion of the gas smooths the actual abundance
gradient is of the order of a few hundred Myrs. Part of the azimuthal
variations currently observed across the Galactic disk might be caused
by changes in the abundance patterns between the spiral arms and the
inter-arm regions ~\cite{kim11a} and by the clumpiness of the star
formation episodes.
However, the kinematics of the above stellar tracers
is quite limited, since they evolve in situ. This working hypothesis is
supported by a very large set of Cepheid abundances provided by
~\cite{luck11a,luck11b}. They found no evidence of azimuthal variations
in an annulus of 1~kpc around the sun.
The statistics concerning the Galactic Cepheids located in the inner
disk is quite limited, and indeed only a handful of Cepheids are known
with Galactocentric distances smaller than 6~kpc. However, in a recent
investigation ~\cite{pedicelli10} by using high-resolution, high
S/N ratio spectra confirmed the super metal-rich nature of four 
of them ~\cite{andrievsky04}.
The hypothesis suggested by ~\cite{davies09a} that the wind of
metal-intermediate bulge stars might mix with metal-rich gas present
along the bar and the Galactic Center to produce a chemical mixture
close to solar appears also very promising. However, to our knowledge
we still lack firm empirical evidence on how the winds of bulge stars
might fall in the Galactic Center. The infall of metal-poor gas in the 
Galactic Center appears an even more plausible channel to explain current 
abundance patterns ~\cite{wakker99,lubowich00}. This is the so-called 
biased infall scenario ~\cite{chiappini99} in which the 
infall of gas takes place more rapidly in the innermost than in the 
outermost regions (inside-out disk formation). 

The abundance pattern of $\alpha$-elements is even more puzzling,
since accurate measurements indicate solar abundances, and therefore
consistent with typical thin disk stars ~\cite{davies09b}.
However, different tracers (B-type stars, red supergiants, classical
Cepheids, HII regions) do provide slightly different mean values,
suggesting a broad distribution (\cite{davies09a}, and references therein).
In this context, it is worth mentioning another piece of evidence concerning
the thin disk. Accurate abundance estimates of $\alpha$-elements in a sizable
sample of classical Cepheids indicates that the $\alpha$(Ca, Mg, Si)-to-iron
ratio attains, within the errors, a constant value across the disk
~\cite{luck06,lemasle07}. This evidence once confirmed
has two relevant implications:

\begin{enumerate}
\item{{\bf Chemical enrichment}-- the chemical enrichment in the last
few hundred Myrs across the Galactic disk seems to be driven by core collapse
supernovae.
}

\item{{\bf Galactic Center}-- the same outcome applies to the Galactic Center,
since current estimates suggest a solar composition for both the iron and the
$\alpha$-elements.
} 
\end{enumerate}

\begin{figure}
\centering
\includegraphics[height=0.50\textheight,width=0.99\textwidth]{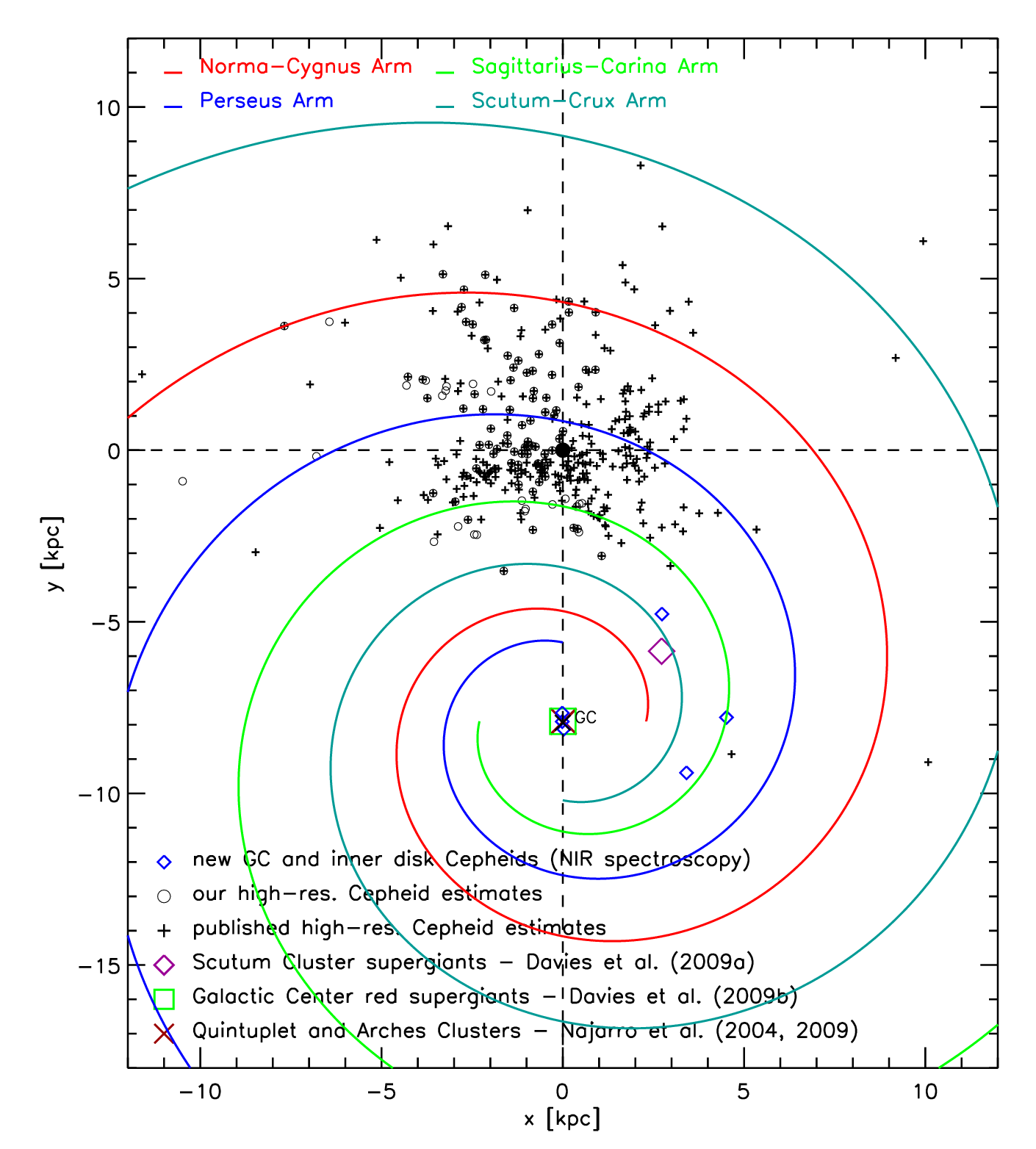}
\caption{Same as the Fig.~3, but projected onto the Galactic plane.
Together with the known classical Cepheids by ~\cite{pedicelli09}
have also been plotted the three classical Cepheids  discovered by ~\cite{matsunaga11}
in the Galactic Center (diamonds) and the three classical Cepheids  recently discovered by
~\cite{matsunaga12} in the inner edge of the thin disk (squares).
The four spiral arms according to the Galactic model of ~\cite{vallee05} 
are plotted with different colors (see labels).
}\label{gal_plane}
\end{figure}
\section{Future perspectives}\label{sec:5}
The content of the previous sections further underline the paramount
effort undertaken by the astronomical community both from the
theoretical and the observational point of view to constrain the
stellar populations, the star formation rate and the chemical
enrichment of the Galactic Center.
During the last few years the multi-wavelength approach also provided a
comprehensive picture of the physical mechanisms driving the formation
and the evolution of disks and bulges ~\cite{kormendy09,graham03}

However, we still lack firm constraints concerning the geometry and the
kinematics at the interface between inner disk, Bulge and Center. The
quoted properties of stellar populations in the Bulge and in the Center
bring forward that old stellar tracers, such as RR Lyrae, appear to be
solid beacons to trace the 3D structure of the innermost Galactic regions.
A detailed map of the RR Lyrae in the Galactic Bulge has been performed
by OGLE ~\cite{soszynski11}, but we still lack a detailed census
of RR Lyrae variables in the Galactic Center.

Moreover, we still lack detailed kinematic constraints concerning 
outer and the inner Lindblad resonance and of the corotation radius 
~\cite{yuan92,yuan97,launhardt02}. 
Recent estimates are mainly based on gas kinematics ~\cite{brand93,honma97,reid09}, 
but estimates based on stellar tracers are 
not available yet. The occurrence of peculiar radial motions at the 
ending points of the Galactic Bar would also provide the optimal target
selection to constrain the chemical enrichment of these regions. 
It is clear that the second generation instruments at ESO/VLT, 
such as the KMOS spectrograph ~\cite{sharples10}, appear 
to be an optimal facility to constrain the kinematics in crowded and highly 
reddened fields. The field of view is slightly larger than 7 arcmin in 
diameter and are available 24 integral field units (IFU). The wavelength 
coverage ranges from 0.8 to 2.5 $\mu$m with a medium spectral resolution
(R$\sim$3500). This new observing facility will provide detailed kinematic 
maps for both cluster and field stars and chemical abundances for a relevant 
fraction of the Galactic Center. 
In this context the next generation of Extremely-Large-Telescopes 
(European-ELT [E-ELT]\footnote{http://www.eso.org/public/teles-instr/e-elt.html},
the Thirty Meter Telescope [TMT]\footnote{http://www.tmt.org/} and the Giant
Magellan Telescope [GMT]\footnote{http://www.gmto.org/}) will also play a key 
role due to their high spatial resolution and high NIR sensitivity. 
The same applies for JWST\footnote{http://ircamera.as.arizona.edu/nircam/}. 

\begin{acknowledgement}
It is a pleasure to thank M. Fabrizio for a detailed and critical reading of 
an early version of this manuscript. One of us GB thanks ASTROMUNDUS and the 
Department of Astronomy, University of Belgrade for their warm hospitality. 
We are indebted to the great patience and the constant support from the 
editors. This work was partially supported by the PRIN~MIUR 2011 (P.I. M. Marconi). 
\end{acknowledgement}

\end{document}